\documentclass[3p]{elsarticle} 
\usepackage[hyphens]{url}

\usepackage{lineno} 

\biboptions{sort&compress} 
\usepackage{graphicx}
\usepackage{booktabs} 

\usepackage[T1]{fontenc}
\usepackage{lmodern}
\usepackage{amssymb,amsmath}
\usepackage{ifxetex,ifluatex}
\usepackage{fixltx2e} 
\IfFileExists{upquote.sty}{\usepackage{upquote}}{}
\ifnum 0\ifxetex 1\fi\ifluatex 1\fi=0 
  \usepackage[utf8]{inputenc}
\else 
  \usepackage{fontspec}
  \ifxetex
    \usepackage{xltxtra,xunicode}
  \fi
  \defaultfontfeatures{Mapping=tex-text,Scale=MatchLowercase}
  
\fi
\IfFileExists{microtype.sty}{\usepackage{microtype}}{}
\usepackage{graphicx}
\makeatletter
\def\maxwidth{\ifdim\Gin@nat@width>\linewidth\linewidth
\else\Gin@nat@width\fi}
\makeatother
\let\Oldincludegraphics\includegraphics
\renewcommand{\includegraphics}[1]{\Oldincludegraphics[width=\maxwidth]{#1}}
\ifxetex
  \usepackage[setpagesize=false, 
              unicode=false, 
              xetex]{hyperref}
\else
  \usepackage[unicode=true]{hyperref}
\fi
\hypersetup{breaklinks=true,
            bookmarks=true,
            pdfauthor={},
            pdftitle={Resolving the measurement uncertainty paradox in ecological management},
            colorlinks=false,
            urlcolor=blue,
            linkcolor=magenta,
            pdfborder={0 0 0}}
\journal{The American Naturalist}
\usepackage[nomarkers]{endfloat}

\begin{document}
\begin{frontmatter}

  \title{Resolving the measurement uncertainty paradox in ecological management}
    \author[a]{Milad Memarzadeh}

    \author[a, 1]{Carl Boettiger}

      \address[a]{Dept of Environmental Science, Policy, and Management, University of
California Berkeley, Berkeley CA 94720-3114, USA}
  
  \begin{abstract}
  Ecological management and decision-making typically focus on uncertainty
  about the future, but surprisingly little is known about how to account
  for uncertainty of the present: that is, the realities of having only
  partial or imperfect measurements. Our primary paradigms for handling
  decisions under uncertainty -- the precautionary principle and optimal
  control -- have so far given contradictory results. This paradox is best
  illustrated in the example of fisheries management, where many ideas
  that guide thinking about ecological decision making were first
  developed. We find that simplistic optimal control approaches have
  repeatedly concluded that a manager should increase catch quotas when
  faced with greater uncertainty about the fish biomass. Current best
  practices take a more precautionary approach, decreasing catch quotas by
  a fixed amount to account for uncertainty. Using comparisons to both
  simulated and historical catch data, we find that neither approach is
  sufficient to avoid stock collapses under moderate observational
  uncertainty. Using partially observed Markov decision process (POMDP)
  methods, we demonstrate how this paradox arises from flaws in the
  standard theory, which contributes to over-exploitation of fisheries and
  increased probability of economic and ecological collapse. In contrast,
  we find POMDP-based management avoids such over-exploitation while also
  generating higher economic value. These results have significant
  implications for how we handle uncertainty in both fisheries and
  ecological management more generally.
  \end{abstract}
 \end{frontmatter}

Imperfect information is ubiquitous in ecological management and
conservation decision making. While the pressing concerns of global
change have put the spotlight on forecasting, i.e.~the uncertainty of
the future (e.g. Petchey et al. 2015), management decisions must also
contend with uncertainty of the present: How many fish are in the sea
today? What regions harbor the most fragile biodiversity? How do we make
use of a rapidly expanding but opportunistically collected data, such as
in citizen science efforts, which may involve strong sampling biases?
Much of our focus in dealing with uncertainty has been through the lens
of statistical approaches to model fitting. The rise of approaches such
as Hierarchical Bayesian Modeling have given us sophisticated tools to
reflect uncertainty in our data and uncertainty in our knowledge of
mechanisms and parameters in our models (Ellison 2004). But estimating a
model is not the same as making a decision, particularly when a model
leaves us with considerable uncertainty about current and future states.
A model can only provide the rules of the chess board. It is the role of
decision theory to provide us with a strategy.

Decision theory has a long history in the ecological literature (e.g
Schaefer 1954; Walters and Hilborn 1978; Clark 1990; Shea 1998; Polasky
et al. 2011) which has sometimes been overshadowed by recent emphasis on
model fitting. While in principle a model will be updated throughout the
decision process (as we will discuss), it is convenient to think of the
model as ``given,'' estimated, with uncertainty, using the best
available methods. The goal of a decision theory is not maximize some
abstract notion of model fit (least squares, likelihood, etc), but a
more practical notion of utility (happiness, economic value, or any
combination of stakeholder objectives, see Halpern et al. (2013)).
Approaches for decision making under uncertainty in ecological systems
can be divided into two camps: (1) those based upon ``optimal control''
solutions, and usually favored by natural resource economists, and (2)
those based in heuristic methods such as scenario planning, resilience
thinking, and precautionary rules of thumb, more commonly found in both
ecological literature and actual practice (Fischer et al. 2009; Polasky
et al. 2011). While researchers have recognized the need to unify the
transparent and quantitative algorithmic approach of optimal control
with the greater complexity and uncertainty of real ecosystems that is
acknowledged by heuristic methods, computational barriers to doing so
have stymied this progress. We illustrate how the limitations of these
approaches have manifested in the example of fisheries management, and
present a new approach that can combine the reality of measurement
uncertainty and the rigor of optimization to resolve a long-standing
paradox and suggest a more robust approach to management.

The challenge of making decisions under uncertainty is not unique to
fisheries. Optimal control-based decision methods which assume perfect
measurements are applied to a wide range of population and ecosystem
management issues, including fire ecology (Richards et al. 1999),
invasive species (Shea 1998; Blackwood et al. 2010), disease outbreaks
(Shea et al. 2014; Li et al. 2017), and protecting biodiversity (Dee et
al. 2017; Iacona et al. 2017), and in common textbooks and reviews
(Mangel 1985; Clark 1990; Marescot et al. 2013). Yet fisheries
conservation and management has long been both a crucible and proving
ground for the theory of ecological management more generally, including
topics such as adaptive management (Walters and Hilborn 1978),
ecosystem-based management (Levin and Lubchenco 2008) and resilience
thinking (Holling 1973; May 1977), while also giving rise to the
sub-discipline of resource economics (Gordon and Press 1954; Schaefer
1954; Beverton and Holt 1957), thanks to its global relevance, long
history, and readily available data. Despite the generality of the
models and issue of measurement uncertainty considered here, any
application must be grounded in both the history and context of the
particular decision problem. In light of this, we begin with a brief
background on existing decision theory approaches common in fisheries
management.

In this paper, we examine three strategies to managing a natural
population, such as a fishery, under such uncertainty. This problem has
a long history in both theory and practice, where seemingly innocuous
simplifying assumptions have led to a deep paradox in how managers
should respond to uncertainty. The precautionary principle (Kriebel et
al. 2001) would seem to suggest that the greater the uncertainty in our
models, the more cautious we should be in management policies. Yet a
well-known mathematical proof by Reed (1979) established that under
quite general conditions, the optimal strategy under uncertainty is no
different than under no uncertainty at all. Attempts to resolve this
have since only deepened this paradox, arguing that adding additional
sources of uncertainty, such as imperfect measurements, should lead to
optimal harvest that increases with greater uncertainty (e.g. Clark and
Kirkwood 1986; Sethi et al. 2005). Meanwhile, much of actual practice
has (perhaps fortunately, as we shall see) ignored these results in
favor of a more heuristic application of the precautionary principle.
Here, we show that this paradox can be resolved by uncovering previous
shortcuts and tackling the optimal management problem head-on, which we
can solve thanks to the computational efficiency of sophisticated
algorithms developed more recently in the field of robotics. This
approach allows us to reject this paradox largely in favor of the
precautionary principle. Further, our comparisons show that widely
recognized existing approaches in both theory and practice are not by
themselves sufficient to manage a population under substantial
uncertainty over the long term, but our recently borrowed approach from
robotics literature can under identical conditions sustain both higher
biomass and higher economic yields than those approaches. This
conclusion has important implications not only for fisheries but
ecosystem more broadly: First, our results highlight that measurement
error must be accounted for in both the decision process itself, and not
merely in the estimation of model. Second, these results underscore the
opportunity to learn from engineering fields such as robotics when faced
with complex and uncertain decisions that we have previously dismissed
as intractable.

\hypertarget{the-decision-problem}{%
\section{The Decision Problem}\label{the-decision-problem}}

We consider the management problem of setting catch quotas for a marine
fishery in the face of imperfect information about the current stock
size and uncertainty about future recruitment. We seek to determine the
sequence of actions \(h_t\) for \(t \in [1, ... \infty]\) that maximize
the net present value (discounted sum of all future profits) of the
fishery. We will denote the discount rate \(\gamma\). For simplicity we
will once again follow classic theory and assume a fixed price for fish,
(equivalently, measuring our value in units of discounted fish rather
than discounted dollars). Each year, the manager also obtains an
estimate \(y_t\) of the true \(x_t\) stock size subject to some
measurement uncertainty \(\xi_t\)

\begin{equation}
y_t = \xi_t x_t 
\end{equation}

For simplicity, we will assume measurement uncertainty is normally
distributed with standard deviation \(\sigma_m\),
\(\xi \sim \mathcal{N}(1, \sigma_m)\). Given an estimated population
model along with this (uncertain) measurement of the stock size, the
manager must choose the set of a harvest quotas \(\lbrace h_t \rbrace\)
for \(t \in [0, 1, \infty]\) that maximizes expected long term utility
of the stock:

\begin{equation}
\max_{ \lbrace h_t \rbrace } \mathbb{E} \left \{ \sum_{t=0}^{\infty} \gamma^t \cdot U(x_t, h_t) \right \}
\end{equation}

For convenience of presentation, we will assume this utility is simply
equal to the harvest itself.

\begin{equation}
 U(x_t, h_t) = \min(h_t, x_t)
\end{equation}

where the \(\min\) function ensures no additional reward for attempting
to harvest more than the entire stock. In economic terms, this
corresponds to a fixed price per fish caught, with no proportional cost
on effort required to do so, but the results do not depend on this
assumption, as existing analyses have previously established (Reed
1979). Using this form allows a more direct comparison to approaches
such as maximum sustainable yield (MSY) which do not include an economic
model but merely maximize long-term catch. In principle, utility could
also reflect ecosystem services or other intrinsic values for fish left
in the sea and not merely the market for consumption (Halpern et al.
2013).

The solution to this decision problem is conditional on a population
model for the underlying ecological process, describing the probability
that the system moves to state \(x_{t+1}\) given that it was in state
\(x_t\) and the manager selected a harvest \(h_t\), which we discuss
below. Such a model would typically be estimated and parameterized from
available data, often (but not necessarily) using a hierarchical
Bayesian approach that acknowledges the existence of uncertainty in both
measurements and process (e.g.~see Dichmont et al. 2016 for a discussion
of how uncertainties are frequently included in stock assessment
models). However, accounting for measurement uncertainty when going from
data to model does not mean we have also accounted for this uncertainty
when going from model to decision. For example, a harvest decision
policy based upon maximum sustainable yield simply depends on the stock
size which maximizes the expected growth rate of the model. Measurement
uncertainty in model estimation can indeed influence the position of
that maximum, but as we shall see, that strategy is only optimal in the
absence of any uncertainty.

There is one further subtlety arising in this seemingly simple problem
statement that we must address. Given our discrete-time formulation of
the recruitment process, it is necessary to decide if the measurement
\(y_t\) happens before or after harvest \(h_t\): that is do we: measure,
recruit, harvest, or measure, harvest, recruit? Following convention in
the optimal control literature (e.g. Reed 1979; Clark 1990; Sethi et al.
2005), we will assume the latter; measurement occurs immediately before
harvest, followed by recruitment. This assumption has occasionally been
reversed to introduce some additional uncertainty from stochastic
recruitment immediately prior to setting the harvest quota (e.g. Clark
and Kirkwood 1986; Weitzman 2002), which appears to mimic uncertainty in
measurements but does not allow uncertainty to accumulate since the
measurement step still occurs without error.

The optimal strategy must think not only about the next year, but rather
considers the sequence of all future actions \({h_t}\), under all
possible realizations of the underlying model. This strategy must also
acknowledge that the manager will receive new information each year, and
may need to modify actions accordingly. It is this ability of the
decision problem to think ahead that allows it to tolerate short-term
costs (e.g.~reduced harvests in the current season) for larger future
payoffs. When Eq (1) is ignored, such optimal control problems are known
as Markov Decision Processes or MDPs. These are more commonly referred
to in the ecological literature by their solution method of Stochastic
Dynamic Programming (SDP; Marescot et al. 2013), and widely used in both
ecological management and animal behavior (Mangel 1985; Mangel and Clark
1988). The addition of measurement uncertainty in Eq (1) turns this
problem into a Partially Observed Markov Decision Process (POMDP), for
which solution methods have not been widely available for large problems
until more recently. We discuss these decision methods in more detail
below.

\hypertarget{population-model}{%
\subsection{Population model}\label{population-model}}

To facilitate tractability and interpretation, we will focus on the
well-studied Gordon-Schaefer model (Gordon and Press 1954; Schaefer
1954) of logistic population growth (stock recruitment),

\begin{equation}
x_{t+1} = \varepsilon_t (x_t-h_t) r  \left(1 - \frac{(x_t-h_t)}{K}\right) \label{Gordon-Schaefer}
\end{equation}

where \(x_t\) is the current stock size, \(h_t\) the harvest chosen that
year, with parameters \(r\) giving the individual growth rate, and \(K\)
the carrying capacity and \(\varepsilon\) representing stochastic
recruitment in a variable environment. We have written this in terms of
\(x_t - h_t\) to underscore that we assume the convention of observe,
harvest, recruit. We will assume for simplicity
\(\varepsilon \sim \mathcal{N}(1, \sigma_g)\). This and similar models
are widely used in large scale analyses across diverse stocks (e.g.
(Costello et al. 2016; Britten et al. 2017)), and form the basis for
much of bioeconomic theory (Clark 1990). As such, it will be easier to
compare our results against intuition and classic theory, and avoid the
possibility of differences arising only because of some particular
subtle assumption hidden in a more complex model.

It is important to remember that all three decision methods we compare
here will use this same model as given. In the case of numerical
simulations, we will simply pick reasonable fixed parameter values for
the model above. To illustrate the application to empirical data, we
will first estimate these parameters using a hierarchical Markov Chain
Monte Carlo (MCMC), and then use the resulting estimates to drive all
three decision models. As such, the details of how the model is
estimated are immaterial to the differences in performance between these
decision methods. In all cases, the model is considered fixed.
Technically, it would be possible to re-estimate the model parameters
given each new observation during the decision process. Adaptations of
the decision processes we consider here include this possibility for
learning about the model during the decisions. Some include the
possibility of taking knowingly bad actions, like suddenly increasing
harvest rates, to improve learning over parameters. We investigate these
in subsequent work (Memarzadeh and Boettiger 2018). For better
comparison with established theory, here, the model is considered fixed.
In the case of the simulations, the model is known without error (though
still stochastic in nature), such that additional observations cannot
improve the parameter estimates in any event. Despite that obvious
simplification, we shall illustrate how existing decision methods fail
to account for measurement uncertainty appropriately. Even with these
simple models, decisions are complex, and knowing the rules of the game
is not the same as knowing how to win.

\hypertarget{current-theory-and-practice}{%
\section{Current Theory and
Practice}\label{current-theory-and-practice}}

Gordon and Press (1954) and Schaefer (1954) independently showed that
the maximum sustainable yield is achieved by reducing the stock to the
population to the size at which it obtains the maximum growth rate. This
stock size is known as Biomass at Maximum Sustainable Yield,
\(B_{MSY}\). For the Gordon-Schaefer model (and many others), this is
achieved at \(B_{MSY} = K/2\). They observed that fishing at a
\emph{constant yield} (that is, an individual fish mortality, \(F\),
such that harvest is biomass times mortality, \(H = F \cdot B\), and
thus \(F_{MSY} := \tfrac{H_{MSY}}{B_{MSY}}\)) will eventually lead to a
population that converges to the biomass \(B_{MSY}\) and produces the
maximum sustainable harvest, \(H_{MSY} = r K / 4\) in this model. From
here, theory and practice diverge.

This \emph{constant yield} (constant mortality) solution thus
corresponds to an equilibrium analysis which does not solve the
time-dependent optimization problem above. In particular, this maximum
sustainable yield (MSY) strategy will not be optimal whenever the stock
is away from \(B_{MSY}\). Clark (1973) demonstrated (assuming no
uncertainty in measurement or stochasticity in population dynamics) that
the optimal time-dependent strategy is not one of constant yield, but
rather of \emph{constant escapement}. We will return to this approach in
a moment.

\hypertarget{uncertainty-in-practice-msy-tac}{%
\subsection{Uncertainty in Practice: MSY \&
TAC}\label{uncertainty-in-practice-msy-tac}}

Maximum Sustainable Yield (MSY) remains the basis of international law
(including the UN, IWC, IATTC, ICCAT, ICNAF; Mace (2001)) and a familiar
standard of management in terrestrial ecosystems as well as aquatic
(Clark 1990). Critics have for some time observed the limitations of
harvesting at MSY in face of uncertainty in stock sizes and population
dynamics (Larkin 1977; Botsford et al. 1997). Many US fisheries reflect
this uncertainty through a series of adjustments that effectively reduce
the target fishing mortality level to reflect this uncertainty.
Typically, a stock assessment model provides a best estimate of
\(B_{MSY}\) and corresponding mortality \(F_{MSY}\) is used to define
the stock Over-Fishing Limit, (OFL). Based on this, a somewhat lower
level is set as the Allowable Biological Catch (ABC), reflecting
uncertainty in the stock assessment. To reflect possible uncertainty
between reported and actual catch, the ABC may be reduced further to
define the Total Allowable Catch, (TAC), which forms the basic unit of
management for many such fisheries. To reflect this process, our
analysis will also consider policies in which the harvest quota is set
at 80\% of the level expected under the MSY policy:
\(H_{TAC}(t) = 0.8 \cdot F_{MSY} \cdot B_t\), for a biomass estimated at
\(B(t)\) in year \(t\). To distinguish this approach from MSY, we will
refer to this approach as a TAC policy. This more closely represents a
heuristic (e.g. Hilborn 2010) or resilience-based approach (sensu
Fischer et al. 2009) than an optimization-based policy. Importantly,
this approach shares the fundamentally stationary assumptions of an MSY
policy by defining a constant mortality rather than a dynamic policy.
Thus, despite being more cautious overall and generating lower economic
yield than a constant escapement policy, TAC policies continue to
harvest at non-zero rates even if a stock falls below \(B_{MSY}\), while
the constant escapement policy does not.

It is important to note that actual management strategies are quite
diverse. Our focus on comparisons TAC, MSY, and Constant Escapement (CE)
reflects only the subset of current practices. Another increasingly
common approach to address uncertainty in both process and models is
known as Management Strategy Evaluation (MSE, Smith 1994; Dorner et al.
2009; Bunnefeld et al. 2011; Punt et al. 2016). MSE performs forward
simulations under a suite of pre-selected candidate strategies to
determine which strategy achieves the best expected outcomes. Crucially,
by limiting the analysis to a suite of candidate strategies over a
finite number of replicate simulations, this approach is able to
accommodate substantially greater complexity and uncertainty than has
been possible with more formal optimal control approaches we discuss
below. Optimal control solutions rely on dynamic programming to consider
all possible strategies (all possible sequences of harvest under all
possible realizations of stock dynamics). As such, MSE can be viewed as
an approximation to the optimal control solutions considered below, a
notion that has been formalized in the literature on approximate dynamic
programming (Nicol and Chadès 2011). Future work may be able to leverage
the forward-simulation approach used in MSE to extend the POMDP methods
introduced here to yet more complex dynamics.

\hypertarget{optimal-management}{%
\subsection{Optimal management}\label{optimal-management}}

So far, attempts to provide a more formal basis for managing uncertainty
than the heuristic adjustment of catch limits described above have
largely foundered in a series of paradoxes. Without uncertainty, the
theoretical policies are quite intuitive: The result derived by Clark
(1973) for an optimal dynamic strategy to replace the equilibrium
solution of MSY can be summarized as: If the stock is most productive at
\(B_{MSY}\), then obtain \(B_{MSY}\) as quickly as possible. Thus, at
any level below \(B_{MSY}\), the economically optimal thing to do is to
completely shut down the fishery, while above this stock, harvests
greater than \(H_{MSY}\) will be needed to bring the stock back to
\(B_{MSY}\). (This intuition must be adjusted slightly in the case of
economic discounting of future profits, but is otherwise quite general,
see Clark (1990)). This strategy is known as \emph{constant escapement}
(CE), since a constant sock size \(B_{MSY}\) escapes harvest each year.
Because CE and MSY converge to the \emph{same long-term biomass and same
long-term average yield}, CE can appear to be merely a more formal
justification for MSY. However, in practice the two approaches are very
different: because natural populations fluctuate year-to-year, under CE,
a different harvest quota must be set each year to achieve the constant
escapement. Because CE is derived as the economic optimum (assuming
perfect measurements), it is sometimes referred to as Maximum Economic
Yield, (MEY; Burgess et al. 2018), or as Rights Based Fisheries
Management (RBFM, Costello et al. 2016).

In the face of a naturally fluctuating population, previous arguments
made by Gordon and Press (1954); Schaefer (1954) or Clark (1973) no
longer hold. As noted above, the optimal harvest of a population under a
model such as the stochastic Gordon Schaefer, Eq
\eqref{Gordon-Schaefer}, is an example of a Markov Decision Problem
(MDP) which must be solved with stochastic dynamic programming. Thanks
to a mathematical proof provided by Reed (1979), fisheries managers have
largely been able to bi-pass such computational effort. However, this
convenient result also opened the door to paradox of uncertainty that
has persisted in ecological management for the past four decades.

\hypertarget{reeds-paradox-s-d}{%
\subsection{\texorpdfstring{Reed's Paradox:
\(S = D\)}{Reed's Paradox: S = D}}\label{reeds-paradox-s-d}}

Reed (1979) proved that, under sufficiently general assumptions, the
optimal escapement \(S\) for a population under stochastic growth, is
identical to Clark (1973)'s optimal escapement \(D\) for a deterministic
population, \(S = D = B_{MSY}\). This surprising result suggests that in
going from a world where a manager has perfect knowledge of absolutely
everything into a scenario where the manager faces considerable
uncertainty about the future state of the world, no additional
precaution is needed.

\hypertarget{clarks-paradox}{%
\subsection{Clark's Paradox}\label{clarks-paradox}}

Clark and Kirkwood (1986) was among the first attempts to resolve Reed's
Paradox. Clark and Kirkwood (1986) (quite correctly, as we will see),
identified the crux of Reed's Paradox as the absence of measurement
uncertainty:

\begin{quote}
An important tacit assumption in Reed's analysis, as in the other works
referred to above, is that the recruitment level X is known accurately
prior to the harvest decision, {[}\ldots{}{]} In the case of fishery
resources, the stock level X is almost never known very accurately,
owing to the difficulty of observing fish in their natural environment.
\end{quote}

Clark and Kirkwood (1986) were unable to solve the resulting problem
exactly, but had to adopt an almost equally troublesome assumption:

\begin{quote}
For reasons of tractability, we shall adopt the simplifying assumption
that the escapement level S, is known exactly at the end of that period.
(The mathematical difficulty of the problem increases markedly if this
assumption is relaxed.)
\end{quote}

Unfortunately, this ``simplifying assumption'' serves to squash most of
the measurement error, and their results instead only deepened the
paradox, finding policies that become even \emph{less} cautious as
uncertainty increases:

\begin{quote}
{[}Our{]} results appear to contradict the conventional wisdom of
renewable resource management, under which high uncertainty would call
for increased caution in the setting of quotas.
\end{quote}

Relying on a quite different but still flawed assumption nearly two
decades later, Sethi et al. (2005) largely confirm Clark's Paradox,
which they likewise observed with some concern:

\begin{quote}
It may seem counter-intuitive that a measurement error causes lower
expected escapements below the deterministic fishery closure threshold.
\end{quote}

Despite these notes of caution, both Clark and Kirkwood (1986) and Sethi
et al. (2005) ultimately attempt to rationalize this counter-intuitive
conclusion rather than reject it. Intuitively, uncertainty in the
population growth model (stochasticity) or measurement error, would seem
to justify harvesting fewer fish. Instead, existing optimal control
theory has presented only the paradoxical results that uncertainty
either should not matter at all, or that we should increase rather than
decrease harvests in response to increased uncertainty. Numerous other
attempts have been made to resolve this paradox through alternative
assumptions, the most common being to assume from the outset to look
only for `constant-escapement' type policies (Ludwig and Walters 1981;
Roughgarden and Smith 1996; Engen et al. 1997; Moxnes 2003). Owing to
the ``mathematical difficulty'' Clark first observed, none have
attempted a direct solution which we will present here.

In practice, TAC-managed fisheries are more common that CE-managed
fisheries, at least in the United States. CE-based regulation is limited
primarily to salmon, species for which population estimates may be more
precise and thus closer to Reed's assumption. We are not aware of any
fishery that has put Clark's Paradox into practice, harvesting more in
response to increased uncertainty. As we shall illustrate, this is
fortunate; though approaches that properly account for both forms of
uncertainty can perform even better.

\hypertarget{pomdps-an-optimal-treatment-of-measurement-uncertainty}{%
\subsection{POMDPs: An optimal treatment of measurement
uncertainty}\label{pomdps-an-optimal-treatment-of-measurement-uncertainty}}

Why does the introduction of measurement uncertainty make the decision
problem so much harder? Consider the thought experiment of managing a
fish stock where in each of the previous four years stock assessments
have put the population size at, say, 51, 54, 49, 51, (say, measured in
percent of carrying capacity, \(K\)). Under a constant escapement target
of 50, the harvest quota would be 1, 4, 0, 1, respectively. Note that
this control rule depends only on the measurement for the year in
question, which is assumed to be measured without error. If we then
measure a stock size of 75, an assumption that this measurement is made
without error would justify the sudden increase to a quota of 25. Yet if
we admit the possibility of measurement error, this sudden bump starts
to look very suspicious in light of all the previous observations.
Perfect measurements let us exploit the Markov property of the model --
the most recent measurement tells us all we need to know. In contrast,
the introduction of measurement uncertainty breaks this Markovian
assumption: a rational decision maker would suspect this 75 to be an
over-estimate based on prior observations. Under imperfect observations
introduced by Eq (1), we no longer have a Markov Decision Process (MDP)
over the observed state space. Rather, we have a Partially Observed
Markov Decision Process (POMDP). A POMDP policy cannot be expressed as a
policy function at all (much less as policy function with a single
constant-escapement target, as much previous work has assumed). Instead,
a POMDP policy can be thought of as depending not only on the most
recent observation, but the recent observation combined with a
\emph{prior belief} about what state the system is in, which itself is
determined by all past observations. In this manner, the POMDP can
consider how much weight to give the most recent observation based on
how it compares to prior beliefs.

Algorithmic innovations from the robotics community have now made it
possible to numerically solve non-trivial POMDP problems. The need to
consider all past observations becomes particularly demanding given the
forward-looking nature of a decision problem. Not only must we consider
all possible sequences of future actions over all possible sequences of
future states, but we must add to that all possible sequences of future
measurements. The computational complexity involved has historically
limited POMDP applications to contexts with only a handful of possible
states and actions (e.g. Chadès et al. 2008, 2011; Fackler and Haight
2014; Fackler and Pacifici 2014), insufficient to capture fisheries
models appropriately. Meanwhile, methods to solve decision problems with
imperfect information have advanced steadily in the field of robotics
and artificial intelligence. Almost contemporaneous with Reed's work,
papers by Smallwood and Sondik (1973) and Sondik (1978) laid
mathematical foundations for algorithms that could efficiently solve
POMDP problems (e.g. Kaelbling et al. 1998; Pineau et al. 2003). More
recently, driven by demands in areas such as autonomous vehicle decision
making, newer point-based approximation algorithms such as SARSOP
(Kurniawati et al. 2008) have made it possible to solve POMDPs with 100s
of states and actions. We have adapted this algorithm for application to
the fisheries context, and provide an efficient implementation as a
preliminary R package (Boettiger et al. 2018) which can be used to
replicate and further explore results presented here. Details of the
analysis presented here including annotated code to reproduce and
further explore all of the following results are provided in the
supplementary material.

Precise definitions of both the (fully observed) MDPs and POMDPs (not to
be confused with Markov Process Models and Hidden Markov Process Models
-- methods for which estimate the model but do not involve the feedback
of any decision process) in an ecological context can be found in
Williams (2011), or in classic work from the engineering and robotics
community (e.g. Smallwood and Sondik 1973; Sondik 1978). For
convenience, we summarize the POMDP definition and common notation here.
The POMDP problem is posed identically to that of the MDP problem, with
the addition of an observation process The POMDP problem for fisheries
question considered here can be summarized as follows:

\textbf{Transition} process (state equation): \(T(x_t, x_{t+1}, a_t)\):
the probability that a system is in state \(x_{t+1}\) at time \(t+1\)
given that it began in state \(x_t\) at time \(t\) and the manager took
action \(a_t\). In our context, this relationship is given by the
Gordon-Schaefer stock recruitment function \(f\) with normally
distributed growth uncertainty
\(x_{t+1} \sim \mathcal{N}(f(x_t,a_t), \sigma_g)\), truncated at zero to
exclude negative population sizes. The action \(a_t\) represents the
harvest quota set for the fishery. Attempting to implement a quota the
exceeds the true population size results in collapse of the fishery. As
Clark and Kirkwood (1986) eloquently argues, assuming that extinction is
impossible regardless of fishing intensity would unreasonably bias the
decision problem.

\textbf{Observation} function: \(O(x_t,y_t,a_{t-1})\) the probability of
observing state \(y_t\) given a system in state \(x_t\). In principle,
the action chosen can influence the precision of the observation. In our
case, we simply assume normally distributed errors around the true
state, \(y_t \sim \mathcal{N}(x_t, \sigma_m)\), truncated at zero to
exclude negative population sizes.

\textbf{Utility} function: \(U(x_t,a_t)\), the utility received at time
\(t\) for taking action \(a_t\), given that the system is in state
\(x_t\). For simplicity of analysis, we will simply set the utility to
be equal to the harvested stock: \(U(x_t, a_t) = \min(x_t, a_t)\),
indicating that realized harvest cannot be negative. This choice ensures
that in the case of no uncertainty (\(\sigma_g = \sigma_m = 0\)), the
optimal solution matches that expected under a simple MSY calculation.
More realistic utility functions may include diminishing returns with
increasing harvest (supply and demand effects), and the cost of fishing,
both of which act to suppress large harvests. By focusing on a simple
utility we can be sure that our comparison to MSY is driven by the
treatment of uncertainty rather than merely differing economic
assumptions.

The optimization problem is to select the action \(a_t\) that will
maximize the net present utility over all time. Future utility may be
discounted by a factor \(\gamma\), so that a value \(V\) in \(t\) year
is valued at \(\gamma^t V\) today. Numerically, each of these functions
are defined over a discrete set of possible states, observations, and
actions, and can thus be represented as a collection of matrices or
tensors.

\hypertarget{simulations}{%
\subsection{Simulations}\label{simulations}}

We consider the average fish biomass across 100 replicate simulations of
the same stock dynamics (Eq 4) under three different management
strategies: constant escapement (CE), total allowable catch (TAC; equal
to 80\% MSY), and partially observed Markov decision process (POMDP)
management. Each suite of simulations is considered under three
sequentially higher measurement error regimes, from \(\sigma_m = 0\) (no
measurement errors), \(\sigma_m = 0.1\) (low error), and
\(\sigma_m = 0.15\) (moderate error), as indicated, while stochastic
recruitment (environmental noise) is set to a moderate
\(\sigma_g = 0.15\). Empirical estimates of measurement error in stock
sizes vary widely, with estimates often dependent upon methodological
assumptions. Early estimates have put possible error as high as 50\%
(Clark and Kirkwood 1986) and more recent analyses have suggested an
average of closer to 36\%, though with cases that may be much higher
(Ralston et al. 2011). Consequently, range between 0 to 15\% provides a
probably conservative estimate of typical uncertainty ranges, which we
show is nevertheless large enough to drive substantially negative
ecological and economic outcomes under management strategies that do not
appropriately account for that error over the long run.

\hypertarget{results}{%
\section{Results}\label{results}}

\begin{figure}
\centering
\includegraphics{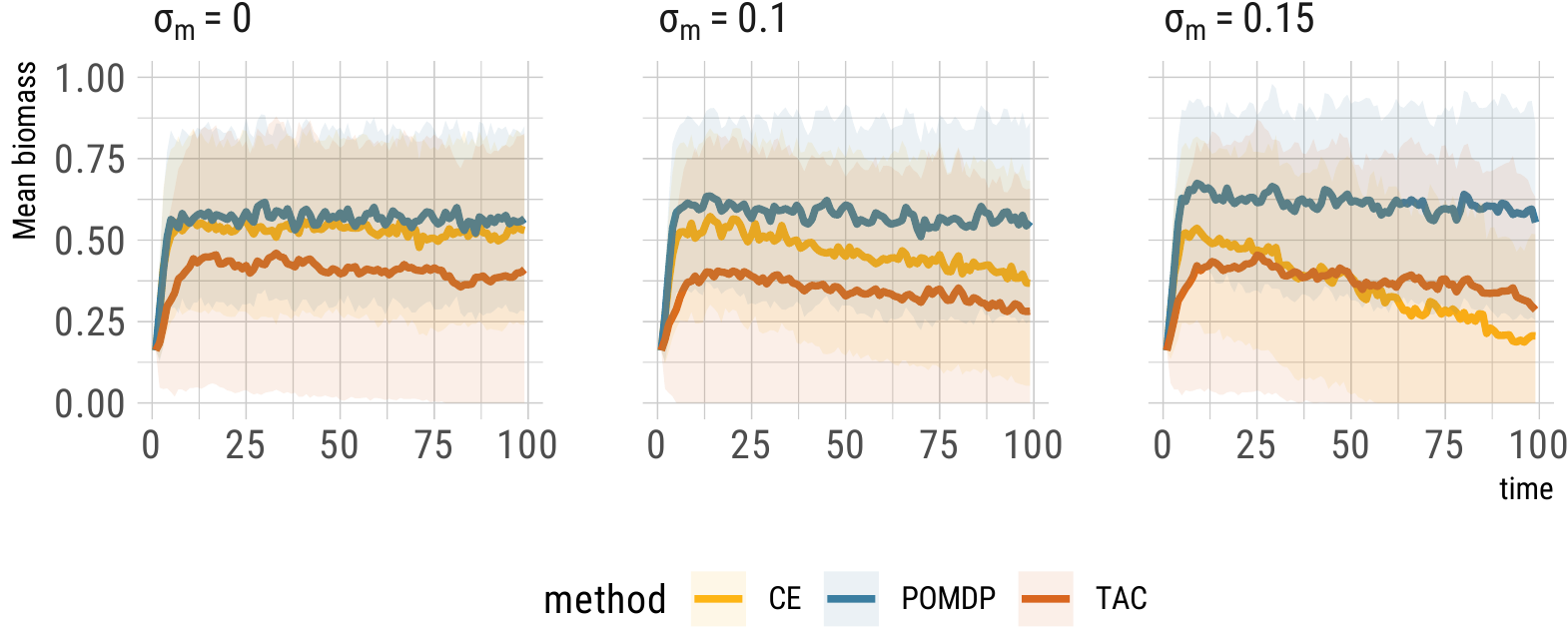}
\caption{Average fish biomass under different management strategies
under increasing levels of measurement uncertainty. Each plot the mean
stock size over time across 100 replicate simulations under each policy:
constant escapement (CE), Total allowable catch (TAC = 80\% MSY), and
the proposed partially observed Markov decision process (POMDP) method.
Standard deviation from the mean across replicate simulations is shown
as faint colored bands, indicating significant variation due to
stochasticity between individual replicates. Measurement error increases
as a normal distribution with standard deviation 0, 0.1, or 0.15, as
indicated at the top of the panel. Environmental stochasticity is fixed
a standard deviation of 0.15 in each panel. Carrying capacity K
normalized to 1, r = 0.75. Additional environmental noise levels and
comparison to MSY rather than TAC can be found in the supplementary
material. \label{sims}}
\end{figure}

In the absence of either environmental noise or measurement error POMDP,
CE, and MSY would converge to stock at the \(B_{MSY}\), while TAC would
maintain the stock at a slightly higher level (see Supplemental
material, Figure S1). The introduction of stochastic growth has a
significant negative impact on the TAC strategy, with a mean biomass
significantly lower than \(B_{MSY}\) (first panel, Figure \ref{sims};
this impact is even more severe for MSY, see Figure S2). Without
measurement error, the CE and POMDP strategies are nearly identical with
both approximately maintaining the stock at \(B_{MSY}\) despite the
significant environmental stochasticity. As measurement error increases
in the subsequent panels, CE and TAC strategies perform increasingly
poorly, while the POMDP continues to maintain the average stock close to
\(B_{MSY}\). Notably, CE is significantly more impacted by measurement
uncertainty than TAC, with CE averaging even lower biomass than TAC
under moderate measurement uncertainty. These results confirm that the
precautionary approach represented by the TAC does indeed prove more
robust to the problem of measurement error than the optimization
solution represented by CE, but is less robust to environmental error.
Many of the replicate simulations under CE experience complete stock
extinction under large measurement error, owing to the arbitrarily large
harvests permitted by the ``bang-bang'' nature of the CE policy. In
contrast, the POMDP approach successfully handles both sources of
uncertainty, maintaining higher stocks, near \(B_{MSY}\) level, and
producing the highest yield.

\begin{figure}
\centering
\includegraphics{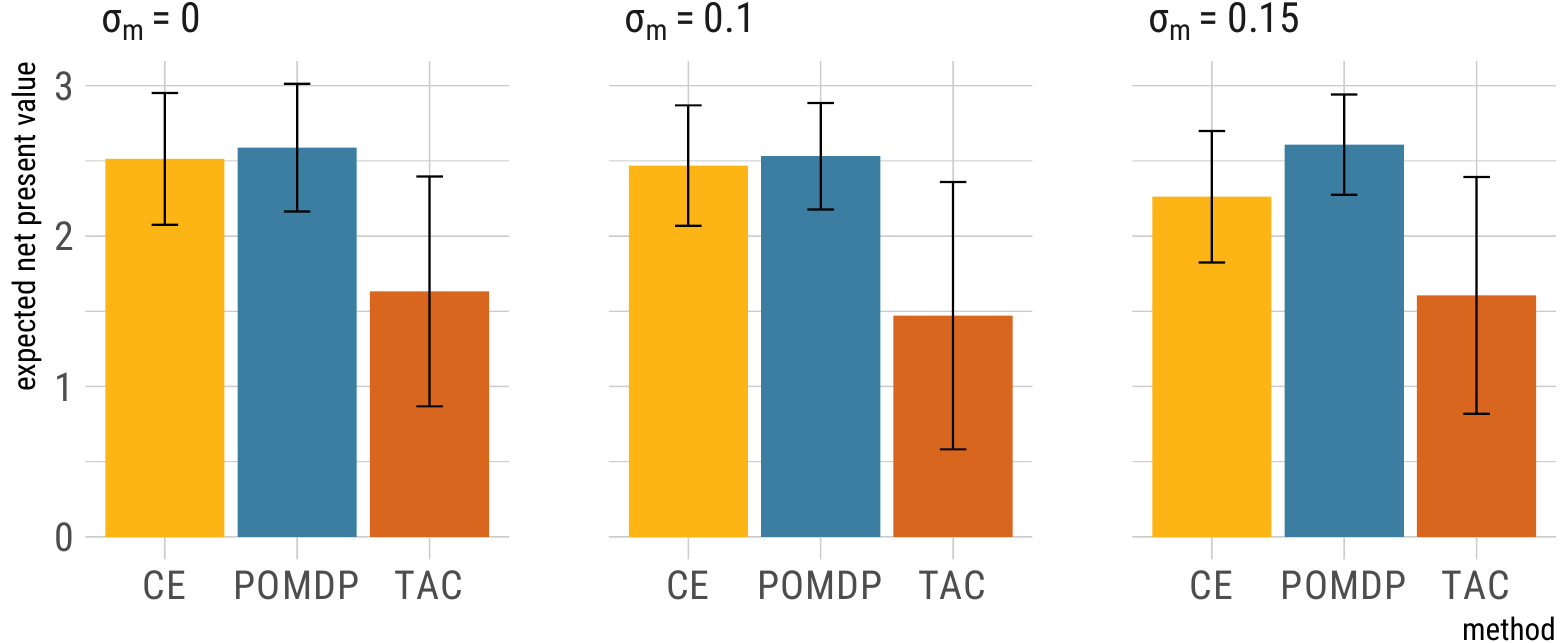}
\caption{Expected net present value of the fishery across strategies
under increasing levels of measurement error \(\sigma_m\). Net present
value is the sum of all future harvests discounted over time and
averaged across 100 replicates simulations under each strategy, assuming
discount factor \(\gamma = 0.95\), as shown in Figure \ref{sims}. The
value under constant escapement (CE) is optimal when measurements are
perfect but decreases rapidly with increasing measurement error. The
more cautious TAC is not economically optimal but largely unimpacted by
increasing error, while POMDP attains consistently high economic yield
despite the increasing uncertainty. \label{econ}}
\end{figure}

This pattern in management success in ecological terms (the relatively
recovery and maintenance of fish biomass) is also borne out in terms of
economic performance. We find the mean net present value (averaging
across replicates and discounting future profits by the discount factor
\(\gamma = 0.95\)) for these same simulations at increasing levels of
measurement uncertainty (Figure \ref{econ}). As before, environmental
stochasticity is set at \(\sigma_g = 0.15\). In the absence of
measurement uncertainty, constant escapement (CE) is optimal (as per
Reed (1979)), while the heuristic caution built into the total allowable
catch (TAC) strategy (at 80\% MSY) results in a sub-optimal economic
yield. The presence of stochasticity in recruitment
(\(\sigma_g = 0.15\)), also contributes to the reduction in yield under
TAC. Though CE is quite robust to the environmental stochasticity level,
this strategy proves very sensitive to increasing measurement error,
showing sharp declines in economic value, falling below the TAC economic
value at \(\sigma_m = 0.15\). Though the risk of stock collapse does
increase with increasing measurement error under TAC (as seen by the
mean declines in Figure \ref{sims}), these have little impact on the
economic value due to the discount rate. A smaller discount rate would
penalize unlikely but not improbable collapses more, since a significant
amount of time is required to realize those rare events. Supplemental
Figure S3 summarizes these economic trends across different noise values
\(\sigma_g\) and includes comparison to a simple MSY policy. In contrast
to the declining economic performance of CE and the consistently
sub-optimal economic yield for TAC, the POMDP strategy continues to
generate economic yields at approximately the optimal level (as attained
by CE in the absence of measurement error) despite the increasingly
uncertain measurements. This demonstrates that the reduced risk of stock
collapse and higher average stock biomass attained by the POMDP strategy
in the presence of high measurement error, Figure \ref{sims} is not the
result of a trivial reduction in harvesting under all circumstances, but
rather, evidence of a more nuanced strategy that manages to account for
the uncertainty in measurement while maintaining a reasonable harvest.

\begin{figure}
\centering
\includegraphics{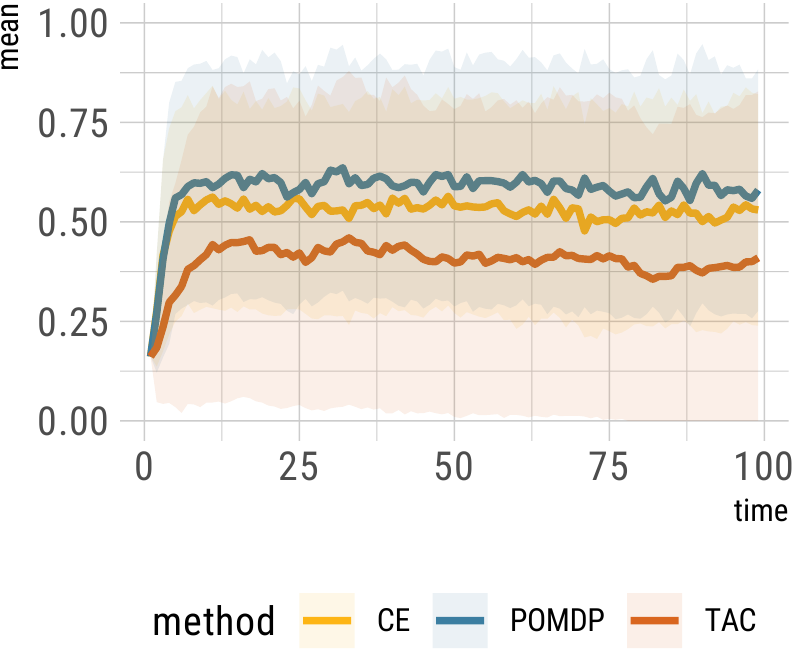}
\caption{Average fish biomass when the POMDP strategy assumes a high
level of measurement uncertainty, while simulations reflect perfect
measurements. Despite this overestimation, biomass under POMDP strategy
closely tracks the optimal biomass under CE in the absence of
measurement error. Solid lines indicate averages over 100 simulations,
colored bands indicate +/- one standard deviation. \label{overest}}
\end{figure}

The POMDP strategy is also robust to overestimation of the level of
measurement uncertainty. In the simulation results shown in Figures
\ref{sims} and \ref{econ}, we have assumed that the level of measurement
uncertainty \(\sigma_m\) was known, and saw that ignoring this
uncertainty (as the CE policy does) has significant negative impact on
ecological and economic outcomes. If the level of measurement
uncertainty is not known precisely, \emph{overestimating} the
measurement error while using the POMDP strategy provides a
precautionary approach that can nevertheless achieve nearly optimal
ecological and economic outcomes. Figure \ref{overest} summarizes the
results of the same simulations as before, but under the scenario in
which the POMDP approach assumes a measurement error of
\(\sigma_m = 0.15\), when in fact all simulated measurements are made
without error (\(\sigma_m = 0\)). This represents an extreme case of
overestimating the measurement error. Stochastic growth remains the same
as before, \(\sigma_g = 0.15\), and simulations under TAC and CE
policies are shown for comparison. When measurement error is absent,
Reed's proofs hold and the CE strategy is optimal. Figure \ref{overest}
shows that the POMDP outcomes track almost exactly the CE outcomes
despite the misplaced assumption that measurements are quite poor. (As
we have already seen, the TAC strategy is insufficiently cautious for
this level of stochasticity in recruitment, resulting in
over-exploitation and long-term decline). The ability of the POMDP
solution to perform nearly optimally even when significantly
overestimating the level of measurement uncertainty contrasts sharply to
the significant declines from ignoring measurement uncertainty seen in
the CE solutions in Figure \ref{sims}. This demonstrates that while we
may not know precisely the level of measurement error, we achieve far
better outcomes overestimating measurement uncertainty than
underestimating it. This also underscores the observation that POMDP
policy is quite robust to the details of the uncertainty. We can get a
better understanding for the performance of POMDP in these simulations
by looking more closely at how any individual decision under a POMDP
strategy compares to the action chosen by the current alternatives.

\begin{figure}
\centering
\includegraphics{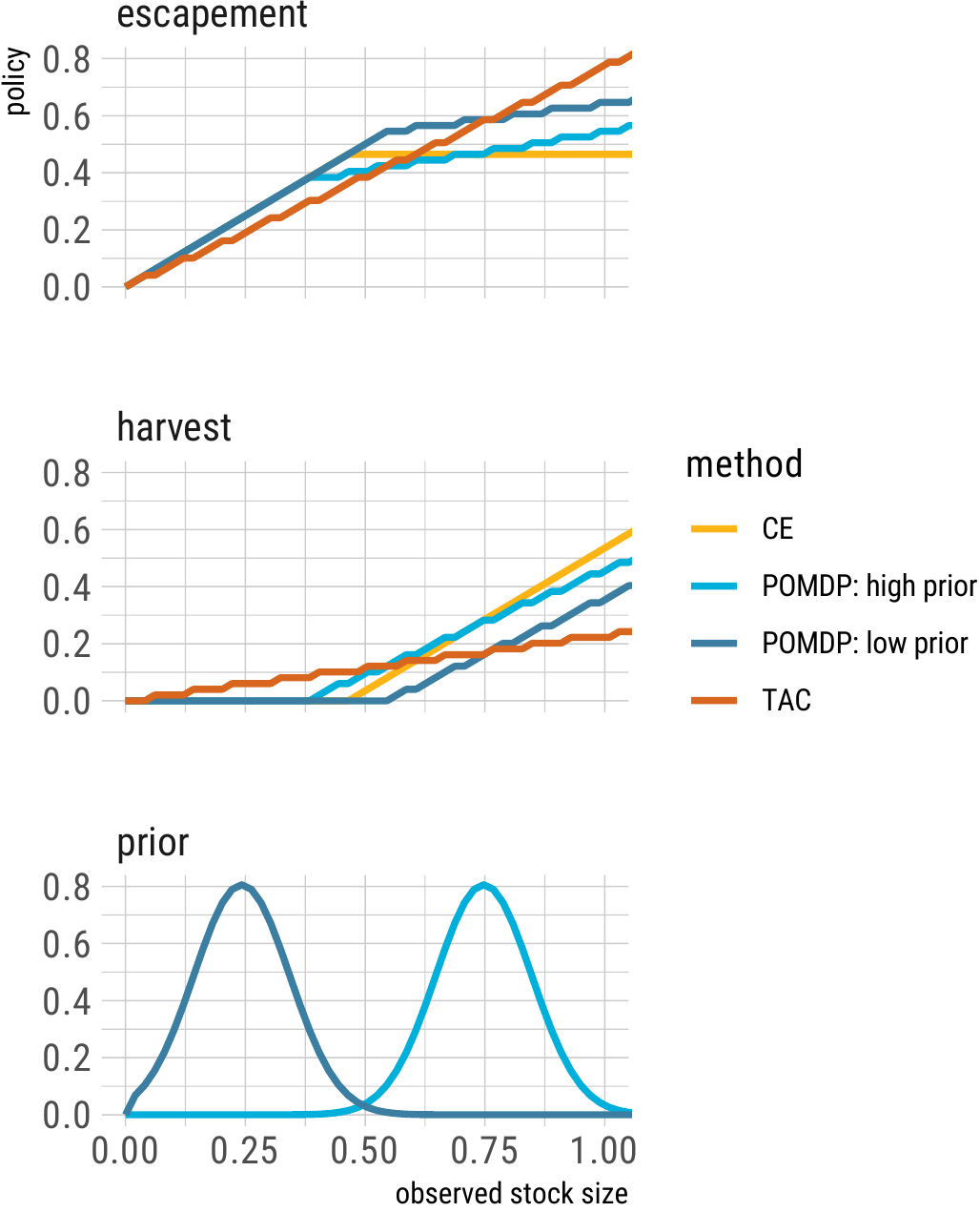}
\caption{Comparison of the harvest and escapement policies under each
strategy as a function of the observed stock size. Escapement refers to
the expected fraction of fish left in the sea, \(x-h\), while harvest
refers to the target catch; two different conventions for plotting the
same action. Uniquely, the POMDP policy plot will depend not only on the
observed stock size, but also depends on prior information (bottom
panel) as determined from any prior observations and actions. Depending
on this prior it may harvest less or more than the other policies given
an identical observation.\label{policy}}
\end{figure}

To better understand the differences in performance of these strategies
and resolve the paradox of uncertainty, we must take a closer look at
how the specific action recommended by each policy compares given the
same observation. Figure \ref{policy} shows the action taken by each
strategy in response to a measurement of the stock size (biomass
estimate). These plots show, for any possible observations, which
strategy will attempt to harvest most and which will attempt to harvest
least. For comparison purposes, we plot policies both in terms of
expected escapement, \(S = Y - H\) as is typical in the optimal control
literature, (e.g. Reed 1979; Clark and Kirkwood 1986; Sethi et al.
2005), and also directly in terms of the harvest quota \(H\). These
plots clearly illustrate the contrast between the constant escapement
(CE) strategy and the precautionary Total Allowable Catch (TAC)
strategy: CE sets harvest strictly to zero for stocks estimated at
biomass below \(B_{MSY}\), while TAC permits a modest harvest of even
very small stocks. (Comparable plots with MSY can be found in the
Supplemental Material). In principle, the CE policy could depend on
\(\sigma_g\), but as Reed (1979) proved, the constant escapement level
with stochasticity is the same as in the deterministic case, S = D,
unless the noise level is quite high (comparison plots in Supplemental
material.) In contrast to this, our POMDP plots depend very much on both
the choice of \(\sigma_m\) and \(\sigma_g\). If \(\sigma_m = 0\), they
reduce exactly to the CE solution. Here we show the corresponding
policies for POMDP solutions focusing on \(\sigma_g = 0.15\) as above,
with a modest measurement uncertainty \(\sigma_m = 0.1\). Alternate
combinations of measurement uncertainty can be found in the appendix but
do not change the general pattern. In addition to an explicit dependence
on the measurement error, our POMDP solutions depend on another piece of
information: any prior observations of the stock size.

Differences in attention to prior beliefs, determined by prior
observations, drive the differences between the POMDP strategy and the
other strategies and can resolve the uncertainty paradox. While the
catch quota under both TAC and CE strategies can be completely
determined given the most recent observation of the stock size by using
the policy curves shown in Figure \ref{policy}, this is not the case for
the POMDP approach. This fundamental difference is key to understanding
the difference in performance and resolving the paradox of uncertainty.
The POMDP policy cannot be specified by the most recent observation
alone. Instead, the POMDP policy depends on all prior observations, not
just the most recent. The reason for this complexity comes from the
Markov property. Observations of the state in the perfectly observed
system satisfy the Markov property: once we have measured the current
biomass exactly, we cannot get any better estimate of the current stock
size by studying older measurements. When measurements are uncertain
this is no longer the case: intuitively, by comparing the most recent
measurement to previous observations we may be able to infer when any
given measurement is unusually high or unusually low. POMDP formalizes
this intuition by capturing the information from all previous
observations into a \emph{prior belief}. This prior belief is updated
after every subsequent action and observation in accordance with Bayes
Law. The mechanics of this process are well documented in the extensive
literature on POMDPs (e.g. Smallwood and Sondik 1973; Sondik 1978;
Kurniawati et al. 2008; Williams 2011), but for our purposes it is
sufficient to observe how this evolving prior belief serves to
continually adjust the POMDP policy. Supplemental Figure S9 (Appendix A)
illustrates how the prior belief evolves in response to subsequent
observations and actions over the course of an individual simulation.

Figure \ref{policy} shows two separate policy curves (in terms of
harvest and escapement) for the same POMDP solution given two different
prior belief distributions (panel 3, priors). While both priors shown
express considerable uncertainty about the precise stock size prior to
the most recent observation, the lower prior is centered at a value
\(\tfrac{1}{4}K\) stock size, while the high prior is centered at a
value of \(\tfrac{3}{4}K\), relative to a (post-harvest, before
observation) target size of \(B_{MSY} = K/2\). The POMDP policy is
determined by the combination of this prior information and the most
recent observation, as indicated by the two different POMDP curves for
harvest/escapement shown corresponding to the different priors. As with
the other policies, the higher the most recent observation (x axis) the
higher the POMDP recommended harvest. Yet unlike the alternative
strategies, the POMDP solution always reflects the prior information.
Consequently, relative to constant escapement (that is, no measurement
error), the POMDP with low prior starts harvesting only at higher stock
sizes and always harvests less. In contrast under the high prior, the
POMDP always harvests at the same or higher level than the constant
escapement solution.

\textbf{A resolution to the paradox.}

Herein lies our resolution to the paradox of uncertainty. Previous work
created this paradox by suggesting that increased harvest rates
(decreased target escapement) would often be the rational response to
increased uncertainty. The exact solutions from the POMDP reveal that
this is only an accident of the assumptions: it is indeed true that
under certain circumstances, harvest levels should increase relative to
the case of no uncertainty, but \emph{only when prior knowledge suggests
the stock size should be much higher than the most recent estimate would
suggest}. In the POMDP solution, all information must be put into its
historical (and constantly updated) context. When a measurement roughly
matches the expectation of this prior context, Figure \ref{policy} shows
that the optimal response from POMDP is roughly comparable or slightly
more cautious to the harvest under no uncertainty, and not more
aggressive as the paradox would suggest. Measurements that exceed
expectations are tempered with some skepticism: while the CE solution is
willing to meet a high stock measurement with a large harvest, the POMDP
solutions increase harvest more cautiously.

This difference between underestimating and overestimating accounts for
the poor performance of the CE solution under large measurement
uncertainty, where it over-harvests whenever measurements are too large.
Even though underestimating is equally likely under the measurement
uncertainty model, sooner or later a run of ``heads'', a sequence of
overestimations relative to the true stock, can drive stocks to very low
levels where the chance of stock extinction becomes possible. It is
precisely this asymmetry: that too much over-harvesting leads to an
irreversible state of extinction, while too much under-harvesting is
always reversible (modulo some lost revenue) that lay behind Clark's
original intuition that there was something fishy about Reed's result
that \(S = D\): that uncertainty required no extra caution. Constant
escapement is particularly susceptible to over-harvesting starting from
stock sizes much higher than \(B_{MSY}\) since its bang-bang optimal
solution attempts to bring stocks back to target level as fast as
possible. CE is not so vulnerable to collapse from small stock sizes,
since it shuts down all harvests once estimates fall below \(B_{MSY}\).
Stock collapse under TAC can also be driven by such a string of heads,
but is unlikely at high stock sizes since harvest never exceeds
\(H_{MSY}\). In both cases, some measurement uncertainty interacts with
inherent stochasticity, which provides a continued source of variation
to sizes above \(B_{MSY}\), where CE strategy is most vulnerable, and
below \(B_{MSY}\), where TAC is most vulnerable.

\begin{figure}
\centering
\includegraphics{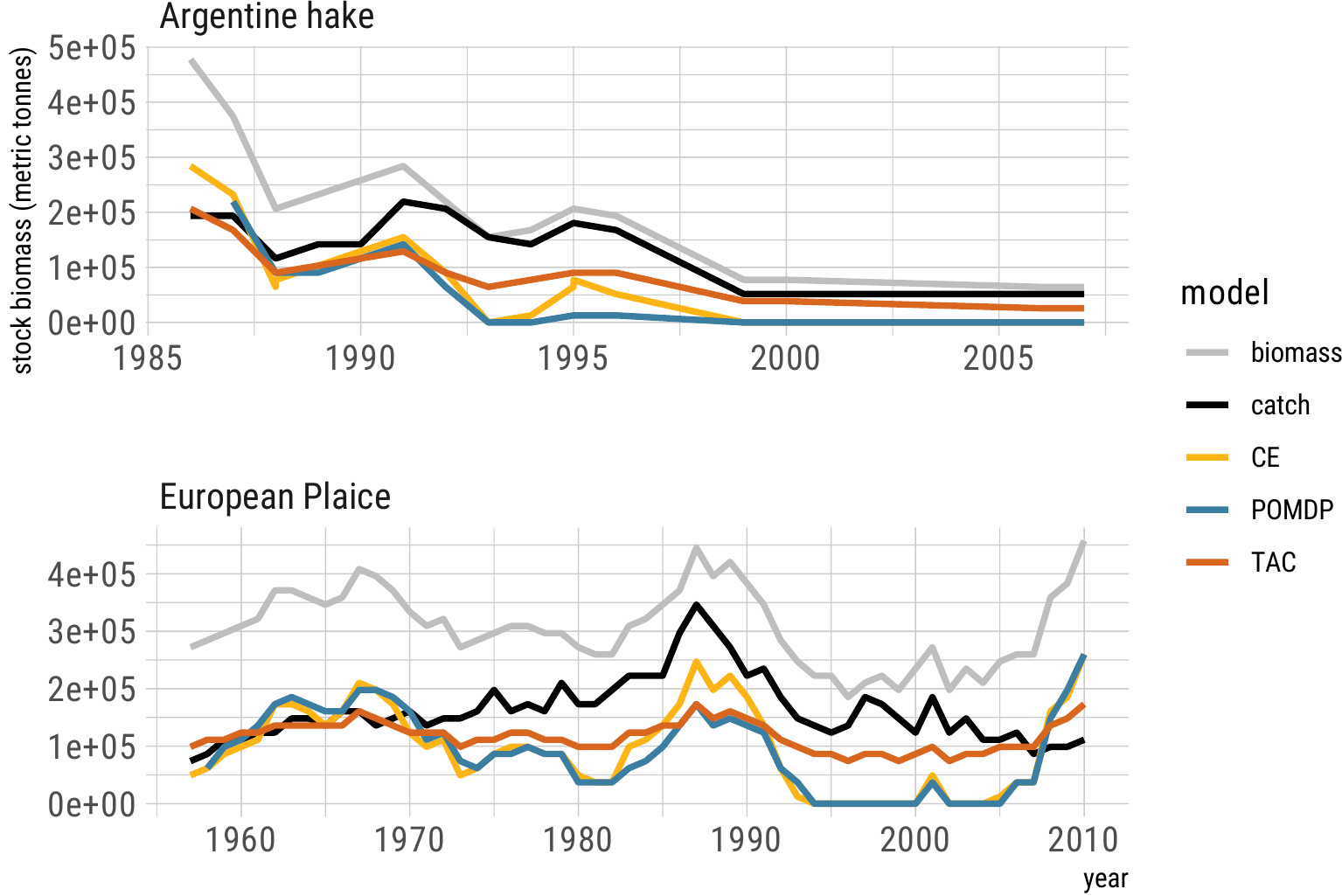}
\caption{Comparisons of harvest level that would be recommended by
policies considered here relative to historical harvest levels in two
commercial fish stocks. POMDP solution assumes a measurement error of
10\%. \label{historical}}
\end{figure}

\textbf{Historical examples and implementation}.

So far we have focused on simulation and an examination of the relative
policies under different prior information. Figure \ref{historical}
compares the harvest level that would have been recommended by each of
the strategies we have considered here against the historically observed
catch recorded for two commercially fished stocks: Argentine Hake and
European Plaice. Historical estimates of biomass and catch are taken
from the R.A. Myers Legacy Stock Assessment database (Ricard et al.
2011). Posterior distributions for parameters for the Gordon-Schaefer
model are estimated with uncertainty through Markov Chain Monte Carlo
using \texttt{nimble} (de Valpine et al. 2017) on the historical data,
illustrating how this process might be done in more complex models as
well (details and code in Supplement, Appendix B). \(B_{MSY}\) and
corresponding TAC and CE policies are calculated based on posterior mean
estimates of the model parameters, along with POMDP solution assuming a
10\% measurement error rate. Though measurement error could be estimated
directly from the raw data, this would not reflect the true measurement
uncertainty arising from the stock assessment process. Each strategy is
then compared to historical observations to determine a recommended
harvest. As we have seen, the TAC and CE harvest policies are uniquely
determined by the observed stock size relative to \(B_{MSY}\), but the
POMDP policy must be re-calculated each time step to reflect both the
prior observation and prior action.

These historical examples provide another useful lens to compare how the
different strategies respond in the face of fluctuations in real data,
rather than model simulations. In both stocks, historical catch almost
always exceeds that recommended by any of our strategies, falling closer
to the MSY value estimated (see Appendix B). In Argentine Hake, the
persistent declines result in both POMDP and CE strategies quickly
closing down the fishery in an attempt to let the stock recover to a
more productive target biomass, while TAC persists with merely a reduced
harvest. A small recovery of biomass in 1995 is met by an immediate
uptick in harvest under CE, while the POMDP response is more
conservative. The example of European Plaice illustrates more volatility
in the stock size, revealing further differences in the strategies.
Despite this volatility, the TAC level remains relatively level across
all five decades, rising or dipping only slightly with changes in stock
size. In contrast, the CE harvest tracks this volatility almost exactly.
Here, the POMDP solution falls in between these extremes, almost
mirroring which-ever of the two policies is more conservative: the POMDP
solution matches the dips in harvest taken by CE to allow the stock to
recover most quickly, but does not track the doubling of harvests
recommended by CE in the mid 1980, increasingly only to the more modest
harvest recommended by the TAC policy. Once again, we see that POMDP
solution provides a more consistently precautionary policy than either
the over-simplified optimal control solution of CE or the more
rule-of-thumb approach represented by TAC. Though we have seen that the
POMDP solution need not always be more conservative (lower harvest) than
these strategies, it makes use of the all prior observations to tune the
level of caution appropriately.

\hypertarget{discussion}{%
\section{Discussion}\label{discussion}}

Using modern algorithms we have been able to both crack the nut of
measurement uncertainty in the optimal management of ecological
populations such as marine fisheries, debunking a long-standing paradox
created by previous literature which has argued that increased
measurement uncertainty should be met with equal or larger harvests
(e.g. Reed 1979; Clark and Kirkwood 1986; Sethi et al. 2005). By using
powerful algorithms developed first in the robotics literature to solve
POMDPs (Kurniawati et al. 2008), we have been able to tackle this
optimal control problem directly rather than resorting to the
simplifying assumptions used in previous work. Our solution confirms the
intuition of the precautionary principle in a quantitatively precise
form: under most (but not all) conditions, \emph{greater uncertainty
results in smaller optimal harvests}. Despite frequently proposing
smaller harvests than existing approaches, the POMDP solution achieves
higher long-term economic value as well as higher average stock sizes
over the long term.

This lesson has clear relevance in the context of fisheries management.
Here, practice is already wiser than theory, where the optimal control
based on the CE results of Clark (1973) and Reed (1979) is limited to
fisheries such as salmon where estimates in stock size may be less
significant than in purely marine species. To our knowledge, no fishery
has intentionally put the more aggressive harvest solutions of Clark and
Kirkwood (1986) and Sethi et al. (2005) into practice. Yet this does not
lift the specter of these methods, where CE-based predictions feature
prominently in global fisheries analyses as the economically optimal
strategy and the fastest route to recovery (Costello et al. 2016;
Burgess et al. 2018). Our results raise serious questions about the
optimistic forecasts predicted under these CE-based policies. Meanwhile,
MSY remains the dominate standard international law (Mace 2001), and
many well-managed US fisheries rely on a version of the TAC based
precautionary adjustment to MSY to comply with catch limits under
Magnuson-Stevens Act. Our results show that these strategies are less
sensitive to measurement error, but not immune. Comparisons against more
heuristic decision methods are more difficult: management strategy
evaluation (MSE; Smith 1994; Punt et al. 2016) will depend on precisely
what management scenarios are considered. Under sufficient uncertainty,
both the existing optimal and precautionary approaches can lead to
declines in biomass and economic returns which could be avoided by
POMDP-based management.

This conclusion has important broader implications for ecological
management more generally. Decision making under uncertainty is a
central challenge for ecology and conservation biology, which frequently
pits optimal control based approaches against more heuristic approaches
such as resilience thinking, scenario planning, and the precautionary
principle (Polasky et al. 2011). With economists frequently favoring the
precise, stylized approach of optimal control and ecologists embracing
the realities of greater complexity and uncertainty echoed in more
heuristic approaches, this division can sometimes appear to pit economic
outcomes against ecological objectives. That perspective is echoed in
the paradox discussed here, in which economic optimization ignoring
uncertainty has argued in cold mathematical logic for a rejection of the
precautionary principle. Our results demonstrate that this divide is
artificial. With sufficiently powerful algorithms we can go some ways to
realizing the call of Fischer et al. (2009) to combine more ecologically
realistic assumptions such as measurement error with optimal control
approaches in ecological management. Both sides are vindicated in our
solution: economic optimization under POMDP does not contradict the
precautionary principle, while under POMDP, optimal control can also
achieve better ecological outcomes than blunt precautionary rule
(i.e.~TAC, which reduced all harvests by 20\% below MSY target). As
Fischer et al. (2009) predicted, unifying optimal control with more
resilience-inspired approaches is win-win.

If accounting for measurement uncertainty can resolve previous paradoxes
and divides, it also raises additional challenges. The approach
discussed here is vulnerable both to the charge of being too simple to
be realistic while also being too complicated to be feasible. These
issues are as easily recognized in other ecological management contexts,
from harvests of wild game or forestry to pest outbreaks, (e.g. Richards
et al. 1999; Shea 1998; Blackwood et al. 2010; Shea et al. 2014; Dee et
al. 2017; Iacona et al. 2017; Li et al. 2017) as in fisheries, where
ecological complexity is invariably high and management capacity
invariably limited. We address each of these in turn.

Are the dynamics considered here too simple? Our underlying model is
undoubtedly much simpler than many of the age or stage structured models
used in modern fisheries management -- aspects whose importance to
ecological dynamics have been established well beyond fisheries as well.
Yet it is critical to bear in mind that this stylized one-dimensional
model is used only to compare the relative performance of the different
decision strategies considered here (TAC, CE, and POMDP) in order to
illustrate the potential importance of uncertainty in measurement. There
is little reason to believe CE or TAC would perform better under more
complex cases such as age-structured models that only further deviate
from the assumptions under which they were derived (e.g. Holden and
Conrad 2015). POMDP approaches are in no way incompatible with more
complex models, whether that be age structure or even more ambitious
non-parametric representations defined directly from data using Gaussian
Processes or other machine learning techniques (e.g. Boettiger et al.
2015). Simple models have always served as tools for comparison and
intuition, and serve the same illustrative role here.

Is this approach too complex to be feasible? The importance of simple
and effective rules of thumb in conservation management has been well
documented (e.g. Chadès et al. 2011). Sophisticated and computationally
intensive approaches such as POMDP may seem improbable at a time when
many areas of natural resource management, even simple, rule-of-thumb
based methods struggle to take root. There is no doubt that the solution
method needed here is computationally intensive even for the simple
model considered here, and will only become more so under greater
ecological complexity. Indeed, it was precisely such computational
limitations that led to the shortcuts in work such as Clark and Kirkwood
(1986), with the ensuing potentially misleading conclusions. While in
the following decades ecologists largely turned away from
computationally intensive optimization required for even simple models
with stochastic dynamic programming, (while continuing to construct ever
more complicated hierarchical models), engineers have chipped away at
the decision problem to reveal algorithms such as SARSOP that make much
larger dynamic programming problems tractable. We may need even more
sophisticated algorithms to accommodate greater ecological complexity,
yet we believe such numerical obstacles are no excuse for management by
half measures. Rather, our example illustrates how ecologists and
managers can learn from and leverage the tools developed in other
disciplines to better face the challenges of management in a complex and
changing world.

\hypertarget{future-directions}{%
\subsection{Future directions}\label{future-directions}}

Several limitations that have been studied in fully observed (MDP)
optimal decision problems, such as parameter uncertainty (e.g. Ludwig
and Walters 1982), model uncertainty (Williams 2001; Boettiger et al.
2015) and adaptive management (e.g. Walters and Hilborn 1976) remain
largely open challenges for partially observed systems. Future work
could extend this analysis to more complex models, such as those with
age structure, as (Holden and Conrad 2015) does for the fully observed
case. Another limiting assumption common to MDPs and POMDPs is that of
stationary dynamics: that the population dynamics equation itself is not
changing over time. In reality, forces such as climate change and other
forms of environmental variations violate this assumption (Britten et
al. 2017). Direct approaches such as Fackler and Pacifici (2014)'s
adaptation of Mixed Observability MDP (Ong et al. 2010) do not scale to
the number of states and actions considered here. A value of information
(VOI) analysis for POMDP (e.g. Johnson and Williams 2015; Memarzadeh and
Pozzi 2016), could identify when it is worthwhile to actively reduce
measurement error.

\hypertarget{acknowledgements}{%
\subsection{Acknowledgements}\label{acknowledgements}}

The authors wish to thank T. Essington for his insight into fisheries
management, and the constructive feedback of the editor and anonymous
reviewers. CB also acknowledges computational resources from NSF's XSEDE
Jetstream (DEB160003) and Chameleon cloud platforms, as well as the
support by the USDA Hatch project CA-B-INS-0162-H.

\hypertarget{references}{%
\section*{References}\label{references}}
\addcontentsline{toc}{section}{References}

\hypertarget{refs}{}
\leavevmode\hypertarget{ref-Beverton1957}{}%
Beverton, R., and S. Holt. 1957. On the Dynamics of Exploited Fish
Populations. Chapman; Hall, London.

\leavevmode\hypertarget{ref-Blackwood2010}{}%
Blackwood, J. C., A. Hastings, and C. Costello. 2010. Cost-effective
management of invasive species using linear-quadratic control.
Ecological Economics 69:519--527.

\leavevmode\hypertarget{ref-Boettiger2015}{}%
Boettiger, C., M. Mangel, and S. Munch. 2015. Avoiding tipping points in
fisheries management through Gaussian process dynamic programming.
Proceedings of the Royal Society B: Biological Sciences
282:20141631--20141631.

\leavevmode\hypertarget{ref-sarsop-pkg}{}%
Boettiger, C., J. Ooms, and M. Memarzadeh. 2018. sarsop: Approximate
POMDP Planning Software in R, v0.5.0,
\textless{}https://github.com/boettiger-lab/sarsop\textgreater{}.

\leavevmode\hypertarget{ref-Botsford1997}{}%
Botsford, L. W., J. C. Castilla, and C. Peterson. 1997. The management
of fisheries and marine ecosystems. Science 277:509--515.

\leavevmode\hypertarget{ref-Britten2017}{}%
Britten, G. L., M. Dowd, L. Kanary, and B. Worm. 2017. Extended
fisheries recovery timelines in a changing environment. Nature
Communications 8:15325.

\leavevmode\hypertarget{ref-Bunnefeld2011}{}%
Bunnefeld, N., E. Hoshino, and E. Milner-Gulland. 2011. Management
stratefy evaluation: a powerful tool for conservation? Trends in Ecology
and Evolution 26:441--447.

\leavevmode\hypertarget{ref-Burgess2018}{}%
Burgess, M., G. McDermott, B. Owashi, L. Peavey Reeves, T. Clavelle, D.
Ovando, B. Wallace, et al. 2018. Protecting marine mammals, turtles, and
birds by rebuilding global fisheries. Science 359:1255--1258.

\leavevmode\hypertarget{ref-Chades2011}{}%
Chadès, I., T. G. Martin, S. Nicol, M. A. Burgman, H. P. Possingham, and
Y. M. Buckley. 2011. General rules for managing and surveying networks
of pests, diseases, and endangered species. Proceedings of the National
Academy of Sciences 108:8323--8.

\leavevmode\hypertarget{ref-Chades2008}{}%
Chadès, I., E. McDonald-Madden, M. a McCarthy, B. Wintle, M. Linkie, and
H. P. Possingham. 2008. When to stop managing or surveying cryptic
threatened species. Proceedings of the National Academy of Sciences
105:13936--40.

\leavevmode\hypertarget{ref-Clark1973}{}%
Clark, C. W. 1973. Profit maximization and the extinction of animal
species. Journal of Political Economy 81:950--961.

\leavevmode\hypertarget{ref-Clark1990}{}%
Clark, C. W. 1990. Mathematical Bioeconomics: The Optimal Management of
Renewable Resources, 2nd Edition. Wiley-Interscience.

\leavevmode\hypertarget{ref-Clark1986}{}%
Clark, C. W., and G. P. Kirkwood. 1986. On uncertain renewable resource
stocks: Optimal harvest policies and the value of stock surveys. Journal
of Environmental Economics and Management 13:235--244.

\leavevmode\hypertarget{ref-Costello2016}{}%
Costello, C., D. Ovando, T. Clavelle, C. K. Strauss, R. Hilborn, M. C.
Melnychuk, T. A. Branch, et al. 2016. Global fishery prospects under
contrasting management regimes. Proceedings of the National Academy of
Sciences 113:5125--5129.

\leavevmode\hypertarget{ref-Dee2017}{}%
Dee, L. E., M. De Lara, C. Costello, and S. D. Gaines. 2017. To what
extent can ecosystem services motivate protecting biodiversity? Ecology
Letters 20:935--946.

\leavevmode\hypertarget{ref-nimble}{}%
de Valpine, P., D. Turek, C. J. Paciorek, C. Anderson-Bergman, T.
Duncan, and R. Bodik. 2017. Programming With Models: Writing Statistical
Algorithms for General Model Structures With NIMBLE. Journal of
Computational and Graphical Statistics 26:403--413.

\leavevmode\hypertarget{ref-Dichmont2016}{}%
Dichmont, C. M., R. A. Deng, A. E. Punt, J. Brodziak, Y.-J. Chang, J. M.
Cope, J. N. Ianelli, et al. 2016. A review of stock assessment packages
in the united states. Fisheries Research 183:447--460.

\leavevmode\hypertarget{ref-Dorner2009}{}%
Dorner, B., R. M. Peterman, and Z. Su. 2009. Evaluation of performance
of alternative management models of Pacific salmon (Oncorhynchus spp.)
in the presence of climatic change and outcome uncertainty using Monte
Carlo simulations. Canadian Journal of Fisheries and Aquatic Sciences
66:2199--2221.

\leavevmode\hypertarget{ref-Ellison2004}{}%
Ellison, A. M. 2004. Bayesian inference in ecology. Ecology Letters
7:509--520.

\leavevmode\hypertarget{ref-Engen1997}{}%
Engen, S., R. Lande, and B. Sæther. 1997. Harvesting strategies for
fluctuating populations based on uncertain population estimates. Journal
of Theoretical Biology 186:201--212.

\leavevmode\hypertarget{ref-Fackler2014b}{}%
Fackler, P., and R. Haight. 2014. Monitoring as a partially observable
decision problem. Resource and Energy Economics 37:226--241.

\leavevmode\hypertarget{ref-Fackler2014}{}%
Fackler, P., and K. Pacifici. 2014. Addressing structural and
observational uncertainty in resource management. Environmental
Management 133:27--36.

\leavevmode\hypertarget{ref-Fischer2009}{}%
Fischer, J., G. D. Peterson, T. a Gardner, L. J. Gordon, I. Fazey, T.
Elmqvist, A. Felton, et al. 2009. Integrating resilience thinking and
optimisation for conservation. Trends in ecology \& evolution
24:549--54.

\leavevmode\hypertarget{ref-Gordon1954}{}%
Gordon, H. S., and C. Press. 1954. The Economic Theory of a
Common-Property Resource: The Fishery. Journal of Political Economy
62:124--142.

\leavevmode\hypertarget{ref-Halpern2013}{}%
Halpern, B. S., C. J. Klein, C. J. Brown, M. Beger, H. S. Grantham, S.
Mangubhai, M. Ruckelshaus, et al. 2013. Achieving the triple bottom line
in the face of inherent trade-offs among social equity, economic return,
and conservation. Proceedings of the National Academy of Sciences
110:6229--34.

\leavevmode\hypertarget{ref-Hilborn2010}{}%
Hilborn, R. 2010. Pretty Good Yield and exploited fishes. Marine Policy
34:193--196.

\leavevmode\hypertarget{ref-Holden2015}{}%
Holden, M., and J. Conrad. 2015. Optimal escapement in stage-structured
fisheries with environmental stochasticity. Mathematical biosciences
269:76--85.

\leavevmode\hypertarget{ref-Holling1973}{}%
Holling, C. S. 1973. Resilience and Stability of Ecological Systems.
Annual Review of Ecology and Systematics 4:1--23.

\leavevmode\hypertarget{ref-Iacona2017}{}%
Iacona, G. D., H. P. Possingham, and M. Bode. 2017. Waiting can be an
optimal conservation strategy, even in a crisis discipline. Proceedings
of the National Academy of Sciences 114:201702111.

\leavevmode\hypertarget{ref-Johnson2015}{}%
Johnson, F. A., and B. K. Williams. 2015. A Decision-Analytic Approach
to Adaptive Resource Management. Pages 61--84 \emph{in} C. R. Allen and
A. S. Garmestani, eds. Adaptive management of social-ecological systems.
Springer Netherlands, Dordrecht.

\leavevmode\hypertarget{ref-Kaelbling1998}{}%
Kaelbling, L. P., M. L. Littman, and A. R. Cassandra. 1998. Planning and
Acting in Partially Observable Stochastic Domains. Artificial
Intelligence 101:99--134.

\leavevmode\hypertarget{ref-Kriebel2001}{}%
Kriebel, D., J. Tickner, P. Epstein, J. Lemons, R. Levins, E. L.
Loechler, M. Quinn, et al. 2001. The precautionary principle in
environmental science. Environmental health perspectives 109:871--6.

\leavevmode\hypertarget{ref-Kurniawati2008}{}%
Kurniawati, H., D. Hsu, and W. S. Lee. 2008. SARSOP : Efficient
Point-Based POMDP Planning by Approximating Optimally Reachable Belief
Spaces. Proceedings of Robotics: Science and Systems IV.

\leavevmode\hypertarget{ref-Larkin1977}{}%
Larkin, P. a. 1977. An Epitaph for the Concept of Maximum Sustained
Yield. Transactions of the American Fisheries Society 106:1--11.

\leavevmode\hypertarget{ref-Levin2008}{}%
Levin, S. A., and J. Lubchenco. 2008. Resilience, Robustness, and Marine
Ecosystem-based Management. BioScience 58:27--32.

\leavevmode\hypertarget{ref-Li2017}{}%
Li, S.-L., O. N. Bjørnstad, M. J. Ferrari, R. Mummah, M. C. Runge, C. J.
Fonnesbeck, M. J. Tildesley, et al. 2017. Essential information:
Uncertainty and optimal control of Ebola outbreaks. Proceedings of the
National Academy of Sciences 114:5659--5664.

\leavevmode\hypertarget{ref-Ludwig1981}{}%
Ludwig, D., and C. Walters. 1981. Measurement errors and uncertainty in
parameter estimates for stock and recruitment. Journal of Canadian
Fisheries and Aquatic Sciences 38:711--720.

\leavevmode\hypertarget{ref-Ludwig1982}{}%
Ludwig, D., and C. J. Walters. 1982. Optimal harvesting with imprecise
parameter estimates. Ecological Modelling 14:273--292.

\leavevmode\hypertarget{ref-Mace2001}{}%
Mace, P. M. 2001. A new role for MSY in single-species and ecosystem
approaches to fisheries stock assessment and management. Fish and
Fisheries 2:2--32.

\leavevmode\hypertarget{ref-Mangel1985}{}%
Mangel, M. 1985. Decision and control in uncertain resource systems 255.

\leavevmode\hypertarget{ref-Mangel1988}{}%
Mangel, M., and C. W. Clark. 1988. Dynamic Modeling in Behavioral
Ecology. (J. Krebs \& T. Clutton-Brock, eds.). Princeton University
Press, Princeton.

\leavevmode\hypertarget{ref-Marescot2013}{}%
Marescot, L., G. Chapron, I. Chadès, P. L. Fackler, C. Duchamp, E.
Marboutin, and O. Gimenez. 2013. Complex decisions made simple: a primer
on stochastic dynamic programming. Methods in Ecology and Evolution
4:872--884.

\leavevmode\hypertarget{ref-May1977}{}%
May, R. M. 1977. Thresholds and breakpoints in ecosystems with a
multiplicity of stable states. Nature 269:471--477.

\leavevmode\hypertarget{ref-Memarzadeh2018}{}%
Memarzadeh, M., and C. Boettiger. 2018. Adaptive management of
ecological systems under partial observability. Biological Conservation
224:9--15.

\leavevmode\hypertarget{ref-Memarzadeh2016b}{}%
Memarzadeh, M., and M. Pozzi. 2016. Value of information in sequential
decision making: component inspection, permanent monitoring and
system-level scheduling. Reliability Engineering \& System Safety
154:137--151.

\leavevmode\hypertarget{ref-Moxnes2003}{}%
Moxnes, E. 2003. Uncertain measurements of renewable resources:
approximations, harvesting policies and value of accuracy. Journal of
Environmental Economics and Management 45:85--108.

\leavevmode\hypertarget{ref-Nicol2011}{}%
Nicol, S., and I. Chadès. 2011. Beyond stochastic dynamic programming: a
heuristic sampling method for optimizing conservation decisions in very
large state spaces. Methods in Ecology and Evolution 2:221--228.

\leavevmode\hypertarget{ref-Ong2010}{}%
Ong, S., S. Png, D. Hsu, and W. Lee. 2010. Planning under uncertainty
for robotic tasks with mixed observability. The International Journal of
Robotics Research 29:1053--1068.

\leavevmode\hypertarget{ref-Petchy2015}{}%
Petchey, O. L., M. Pontarp, T. M. Massie, S. Kéfi, A. Ozgul, M.
Weilenmann, G. M. Palamara, et al. 2015. The ecological forecast
horizon, and examples of its uses and determinants. Ecology Letters
18:597--611.

\leavevmode\hypertarget{ref-Pineau2003}{}%
Pineau, J., G. Gordon, and S. Thrun. 2003. Point-based value iteration:
An anytime algorithm for POMDPs. IJCAI International Joint Conference on
Artificial Intelligence 1025--1030.

\leavevmode\hypertarget{ref-Polasky2011}{}%
Polasky, S., S. R. Carpenter, C. Folke, and B. Keeler. 2011.
Decision-making under great uncertainty: environmental management in an
era of global change. Trends in Ecology \& Evolution 26:398--404.

\leavevmode\hypertarget{ref-Punt2016}{}%
Punt, A. E., D. S. Butterworth, C. L. de Moor, J. A. A. De Oliveira, and
M. Haddon. 2016. Management strategy evaluation: best practices. Fish
and Fisheries 17:303--334.

\leavevmode\hypertarget{ref-Ralston2011}{}%
Ralston, S., A. E. Punt, O. Hamel, J. D. Devore, and J. R. Conser. 2011.
A meta-analytic approach to quantifying scientific uncertainty in stock
assessments. Fishery Bulletin 217--231.

\leavevmode\hypertarget{ref-Reed1979}{}%
Reed, W. J. 1979. Optimal escapement levels in stochastic and
deterministic harvesting models. Journal of Environmental Economics and
Management 6:350--363.

\leavevmode\hypertarget{ref-RAM}{}%
Ricard, D., C. Minto, O. Jensen, and J. Baum. 2011. Examining the
knowledge base and status of commercially exploited marine species with
the RAM Legacy Stock Assessment Database. Fish and Fisheries
13:380--398.

\leavevmode\hypertarget{ref-Richards1999}{}%
Richards, S. A., H. P. Possingham, and J. Tizard. 1999. Optimal Fire
Management for Maintaining Community Diversity. Ecological Applications
9:880--892.

\leavevmode\hypertarget{ref-Roughgarden1996}{}%
Roughgarden, J. E., and F. Smith. 1996. Why fisheries collapse and what
to do about it. Proceedings of the National Academy of Sciences
93:5078--5083.

\leavevmode\hypertarget{ref-Schaefer1954}{}%
Schaefer, M. B. 1954. Some aspects of the dynamics of populations
important to the management of the commercial marine fisheries. Bulletin
of the Inter-American Tropical Tuna Commission 1:27--56.

\leavevmode\hypertarget{ref-Sethi2005}{}%
Sethi, G., C. Costello, A. Fisher, M. Hanemann, and L. Karp. 2005.
Fishery management under multiple uncertainty. Journal of Environmental
Economics and Management 50:300--318.

\leavevmode\hypertarget{ref-Shea1998}{}%
Shea, K. 1998. Management of populations in conservation, harvesting and
control. Trends in Ecology and Evolution 13:371--375.

\leavevmode\hypertarget{ref-Shea2014}{}%
Shea, K., M. J. Tildesley, M. C. Runge, C. J. Fonnesbeck, and M. J.
Ferrari. 2014. Adaptive Management and the Value of Information:
Learning Via Intervention in Epidemiology. PLoS Biology 12:9--12.

\leavevmode\hypertarget{ref-Smallwood1973}{}%
Smallwood, R. D., and E. J. Sondik. 1973. The Optimal Control of
Partially Observable Markov Processes Over a Finite Horizon. Operations
Research 21:1071--1088.

\leavevmode\hypertarget{ref-Smith1994}{}%
Smith, A. 1994. Management strategy evaluation -- the light on the hill.
Population dynamics for fisheries management 249--253.

\leavevmode\hypertarget{ref-Sondik1978}{}%
Sondik, E. J. 1978. The Optimal Control of Partially Observable Markov
Processes Over the Infinite Horizon : Discounted Costs. Operations
Research 26:282--304.

\leavevmode\hypertarget{ref-Walters1976}{}%
Walters, C. J., and R. Hilborn. 1976. Adaptive Control of Fishing
Systems. Journal of the Fisheries Research Board of Canada 33:145--159.

\leavevmode\hypertarget{ref-Walters1978}{}%
---------. 1978. Ecological Optimization and Adaptive Management. Annual
Review of Ecology and Systematics 9:157--188.

\leavevmode\hypertarget{ref-Weitzman2002}{}%
Weitzman, M. L. 2002. Landing Fees vs Harvest Quotas with Uncertain Fish
Stocks. Journal of Environmental Economics and Management 43:325--338.

\leavevmode\hypertarget{ref-Williams2001}{}%
Williams, B. K. 2001. Uncertainty , learning , and the optimal
management of wildlife. Environmental and Ecological Statistics
8:269--288.

\leavevmode\hypertarget{ref-Williams2011}{}%
---------. 2011. Resolving structural uncertainty in natural resources
management using POMDP approaches. Ecological Modelling 222:1092--1102.

\end{document}